\documentclass{jpp}
\usepackage{graphicx,amsmath,multirow,color}

\newcommand{\be}{\begin{equation}}
\newcommand{\ee}{\end{equation}}
\newcommand{\rmd}{{\mathrm{d}}}
\newcommand{\xpoint}{\mathrm{X}}
\newcommand{\opoint}{\mathrm{O}}
\newcommand{\sx}{S_\xpoint}
\newcommand{\so}{S_\opoint}
\newcommand{\phiox}{\Phi_{\opoint\xpoint}}
\newcommand{\psitz}{\psi_t^0}
\newcommand{\psipz}{\psi_p^0}

\newcommand{\subitalo}{{\mathrm{I}}}
\newcommand{\rhoi}{{\rho_\subitalo}}
\newcommand{\thetai}{{\vartheta_\subitalo}}
\newcommand{\phii}{{\varphi}}
\newcommand{\torfluxi}{{\psi_{\subitalo\,t}}}
\newcommand{\polfluxi}{{\psi_{\subitalo\,p}}}
\newcommand{\voli}{{V_\subitalo}}
\newcommand{\iotai}{{\iota_\subitalo}}
\newcommand{\widthi}{{W_\subitalo}}
\newcommand{\jaci}{J_\subitalo}
\newcommand{\helfluxi}{{\psi_{\subitalo\,\mathrm{H}}}}

\newcommand{\subsheq}{{\mathrm{H}}}
\newcommand{\rhos}{{\psi_\subsheq}}
\newcommand{\thetas}{{\vartheta_\subsheq}}
\newcommand{\phis}{{\varphi}}
\newcommand{\us}{{u_\subsheq}}
\newcommand{\torfluxs}{{\psi_{\subsheq\,t}}}
\newcommand{\polfluxs}{{\psi_{\subsheq\,p}}}
\newcommand{\vols}{{V_\subsheq}}
\newcommand{\iotas}{{\iota_\subsheq}}

\title{Outside and inside a magnetic island: different perspectives to describe the same observables}
\author{B. Momo\aff{1} and I. Predebon\aff{1,2}\corresp{\email{italo.predebon@igi.cnr.it}}}

\affiliation{
  \aff{1}Consorzio RFX (CNR, ENEA, INFN, Universit\`a di Padova, Acciaierie Venete SpA), Padova, Italy
  \aff{2}Istituto per la Scienza e Tecnologia dei Plasmi - CNR, Padova, Italy
}

\date{\today}

\begin{document}

\maketitle


\begin{abstract}
We compare three different approaches to describe a magnetic island in a generic toroidal plasma: (i) perturbative, from the perspective of the equilibrium magnetic field and the related action in a variational principle formulation, (ii) again perturbative, based on the integrability of a system with a single resonant mode and the application of a canonical transformation onto a new island equilibrium system, and (iii) non-perturbative, making use of a full geometric description of the island considered as a stand-alone plasma domain. For the three approaches, we characterize some observables and discuss the respective limits.
\end{abstract}


\section{Introduction}

The tearing instability, growing near the rational surfaces, leads to helical magnetic perturbations that can change the magnetic topology, with the formation of magnetic islands through a reconnection process in which the field lines break and reconnect. Especially for the description of magnetic field line trajectories, it is convenient to express the magnetic field in terms of its vector potential. In this way, the magnetic field line equations can be derived from a variational principle, formally identical to the action principle in phase space with a Hamiltonian $H$ \citep{cary1983, elsasser1986, hazeltine1992}. The magnetic field line equations are the path that extremize the action $S_\gamma$, and are formally identical to the canonical equations of motion in phase space. The identification of canonical and magnetic variables follows the paper by \citet{pina1988}: the symmetry coordinate (e.g. the toroidal angle $\varphi$ in an axisymmetric magnetic field) is identified with the time $t$, another space coordinate (the poloidal angle $\vartheta$) plays the role of the canonical position $q$, whereas the poloidal and toroidal magnetic fluxes, $\psi_p$ and $\psi_t$, have their equivalence in the Hamiltonian $H$ and canonical momentum $p$, respectively (when the symmetry coordinate is the toroidal one). Hidden in these identifications is the additional equivalence between the covariant components $A_i$ of the vector potential and the magnetic fields, that follows from Stokes theorem. A pedagogical presentation of the above elements is available in the recent review paper by \citet{escande24}.

Following the Hamiltonian formulation of the magnetic field line equations, in this work we compare three different methods to characterize a magnetic island in terms of some observables (e.g., the island width or its volume) which cannot depend on a particular coordinate system, vector potential gauge, or choice of perturbative/nonperturbative approach. The first two methods consider the island as a single-mode perturbation of the equilibrium Hamiltonian and provide a description of the observables, say, from outside the island; conversely, the other method considers the island as a stand-alone plasma subdomain with a self-consistent representation of the observables from its inside.

With the first perturbative approach, we consider an $(m,n)$ tearing mode (where $m$ and $n$ are the poloidal and the toroidal mode number, respectively) at the resonant surface where the island opens, defined by the rational value $\iota=n/m$ of the rotational transform profile. We apply a new formulation for the island width based on the definition of the action $S_\gamma$ of the magnetic system that returns the same result as the island width estimated from the amplitude of the eye-of-cut of a pendulum Hamiltonian in phase space \citep{escande24}. The width classically depends on a flux $\Phi$ related to the radial magnetic perturbation at the rational surface \citep{white13,park08}. In \citet{escande24} this flux is identified -- with a clear geometrical meaning and independently of the coordinate system -- as the flux through the ribbon enclosed by the orbits of the $\opoint$-point and the $\xpoint$-point of the island. In this work we extend its validity even to non-perturbative contexts, showing that it can be interpreted as the helical flux through the island separatrix, independently of the approach adopted.

In the second method, we exploit the integrability of the Hamiltonian of a perturbed system that preserves the helical symmetry, defining the island domain as a new equilibrium with its own magnetic axis corresponding to the $\opoint$-point of the island. Magnetic coordinates are defined on the island flux surfaces as canonical action-angle coordinates, providing a definition of magnetic fluxes through the island flux surfaces, as well as other quantities like the island volume and width \citep{martines11, momo11}. The transition from a perturbed Hamiltonian to a new equilibrium Hamiltonian represents the change of perspective claimed in the title of the paper, moving from the view of an axisymmetric equilibrium with an external magnetic axis with respect to the island (perspective from outside the island) to the island domain itself with its own action-angle coordinates (perspective from inside the island).

In the third, non-perturbative approach, the island domain is considered independently of the surrounding plasma. It is geometrically characterized in terms of magnetic coordinates and metric tensor starting from a discretized field map, again providing integral quantities like the magnetic fluxes, the island volume and width, which are, in principle, measurable \citep{predebon18}.

The three methods are compared for the calculation of several observables of two experimental islands, $(1,1)$ for a circular tokamak and $(1,7)$ for a reversed-field pinch (RFP) plasma. The agreement is satisfactory. In particular, the comparison of the island width provides a first validation of the formula introduced by \citet{escande24} in a perturbative context, and shows the validity of the new interpretation of the flux $\Phi$ (hereafter $\phiox$) as the helical flux through the island separatrix, particularly relevant when the island is characterized in non-perturbative contexts.

The paper is structured as follows: the two perturbative approaches are introduced in section \ref{sec:b} and \ref{sec:sheq}, and the non-perturbative approach in section \ref{sec:inside}; in the following section \ref{sec:c} we compare the three methods for a $(1,1)$ tokamak island and a $(1,7)$ reversed-field pinch (RFP) island in a toroidal device with circular cross-section, with a brief summary closing the paper.


\section{Magnetic islands in a perturbative approach}
\label{sec:b}

\begin{figure}
\includegraphics[width=\columnwidth]{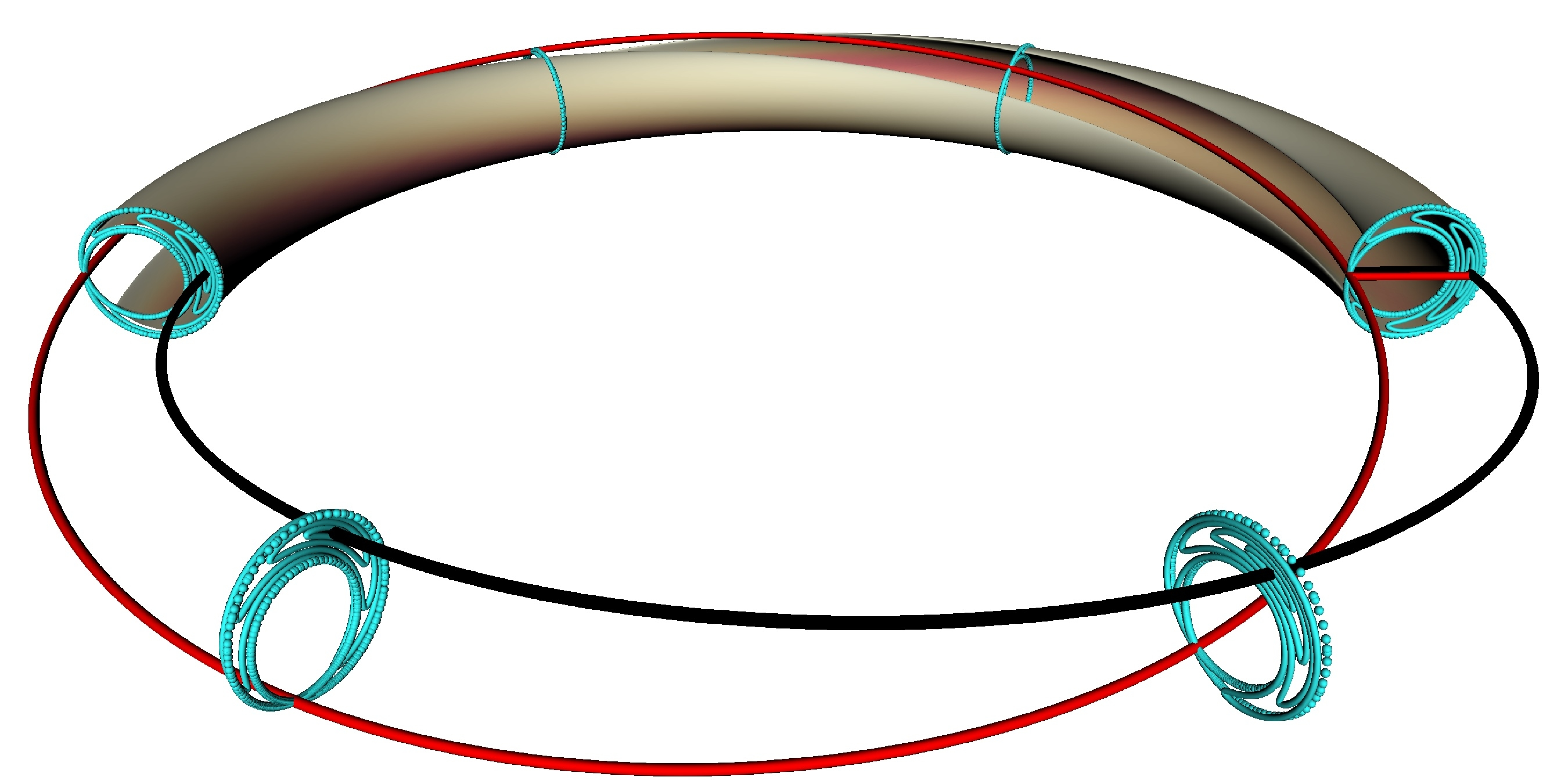}
\caption{Helical ribbon defining the $\phiox$ flux for a $(1,1)$ magnetic island, with in black and red the closed orbits corresponding to the $\opoint$- and the $\xpoint$-point, respectively.}
\label{fig:nastroDE}
\end{figure}

Magnetic islands are due to non-vanishing resonant magnetic perturbations in the plasma, and a perturbative approach is therefore frequently used for their description. In particular, a magnetic island with poloidal $m$ and toroidal $n$ periodicity opens around the resonant flux surface defined by the rational value $\iota=n/m$ of the rotational transform related to the unperturbed equilibrium configuration.

The island width is commonly computed in terms of the geometric width of a pendulum-like eye-of-cat in the context of small resonant perturbations of a regular magnetic field, associated to a time-independent Hamiltonian \citep{hazeltine1992, white13}.
In all cases, the width of a magnetic island results proportional to the square root of a magnetic flux, which turns out to be the perturbation of a helical flux evaluated on the resonant flux surface \citep{park08,predebon16}. As shown in section \ref{sec:sheq}, this flux can be interpreted as the helical flux through the separatrix.

In this section we exploit the existence of a coordinate-independent magnetic flux related to a magnetic island that correctly estimates its width, as proved in \citet{escande24}: this flux, named $\phiox$, is defined for each magnetic island through the ribbon defined by the periodic orbits related to the $\opoint$ and $\xpoint$ points shown in figure \ref{fig:nastroDE}.
The above path is based on the variational principle formulation for magnetic field lines and on the related action
\be
\label{actionB}
S_\gamma = \int_{\gamma} \mathbf{A}(\mathbf{x}) \cdot \rmd \mathbf{x}
\ee
where $\mathbf{x}$ is the spatial coordinates vector, $\mathbf{A}$ is the vector potential, $\mathbf{B} = \nabla \times \mathbf{A}$, and the integral runs along the path $\gamma$ between two points of a magnetic field line. When $\gamma$ is a closed circuit, Stokes theorem states that the action $S_\gamma$ is the magnetic flux through the surface having this circuit as a boundary. 

In order to relate the action $S_\gamma$ to the resonant perturbation that opens an island, the formal identification between canonical and magnetic coordinates, and the equivalence between the covariant components $A_i$, $i=1,2,3$ (where the index $i$ corresponds e.g. to the radial, poloidal and toroidal coordinate, respectively) of the vector potential and the magnetic fluxes must be used in the definition of $S_\gamma$. This equivalence is only valid in the axial gauge $A_1=0$,
\begin{eqnarray}
\label{actionB_A}
S_\gamma &=& \int_{\gamma} (A_2 \,\rmd x^2 + A_3 \,\rmd x^3) \nonumber\\
\label{actionB_p}
&=& \frac{1}{2 \pi} \int_{\gamma} (p \,\rmd \vartheta - H \,\rmd \varphi) \nonumber\\
\label{actionB_fl}
&=& \frac{1}{2 \pi} \int_{\gamma} (\psi_t \,\rmd \vartheta - \psi_p \,\rmd \varphi)
\end{eqnarray}
with $x^2 = \vartheta$ and $x^3 = \varphi$ poloidal and toroidal angles; the relations $p=\psi_t$ and $H=\psi_p$ define the identifications between the canonical momentum and the toroidal magnetic flux and between the Hamiltonian and the poloidal flux, respectively, assuming the canonical position and time to be: $q=x^2=\vartheta$ and $t=x^3=\varphi$. 

Figure \ref{fig:nastroDE} visualizes the helical ribbon defined by the orbits of the $\opoint$ and $\xpoint$ points and helps understand the geometrical meaning of the flux $\phiox$ through that ribbon. Using the definition of the action for a magnetic field line, the $\phiox$ flux turns out to be $\so-\sx$, where $\so$ is the action computed along the closed orbit defined by the $\opoint$ point, and $\sx$ the action along the closed orbit defined by the $\xpoint$ point. In the rest of this section we revisit the derivation of section 5 of the review \citet{escande24} writing the island width as a function of $\phiox$.

Let $ \mathbf{x}= (\psitz,\vartheta,\varphi)$ be magnetic coordinates for the unperturbed equilibrium. If the perturbation is not large enough to violate the requirement of a non-null Jacobian, then the full perturbed system around the resonant surface, in the same $\mathbf{x}$ coordinate system, is approximated by
\begin{eqnarray}
\label{psit_sh}
\psi_t(\mathbf{x}) \simeq \psitz + \psi_t^{m,n}(\psitz) \, e^{i u} + c.c. \\
\label{psip_sh}
\psi_p(\mathbf{x}) \simeq \psipz + \psi_p^{m,n}(\psitz) \, e^{i u}  + c.c.
\end{eqnarray}
where $u=m \vartheta-n \varphi$ is called helical angle and $c.c.$ indicates the complex conjugation.
The unperturbed flux, $\psipz=\psipz(\psitz)$, defines the unperturbed equilibrium and its flux surfaces through the relation $\iota=\rmd \psipz/\rmd \psitz=\rmd \vartheta / \rmd \varphi$. The $\psi^{m,n}(\psitz)=\left|\psi^{m,n}\right| e^{i \alpha^{m,n}}$ terms are the Fourier components of the fluxes having the resonant periodicity. 

We now better define the actions $\so$ and $\sx$, which are the line integrals along the lines defined by the $\opoint$ and $\xpoint$-point of the $(m, n)$ magnetic island, respectively. In calculating $\so$ and $\sx$ from equation \eqref{actionB_fl} using definitions \eqref{psit_sh}-\eqref{psip_sh} we assume that $m$ and $n$ are mutually relatively prime; $\varphi$ varies by $2\pi m$ along $\opoint$ or $\xpoint$ and $\vartheta$ varies by $2\pi n$, while the helical angles along the $\opoint$ and $\xpoint$ orbits ($u_\opoint$ and $u_\xpoint$ respectively) are constant.

We first compute the action $\so$:
\begin{eqnarray}
\label{So_int}
\so &=& \frac{1}{2 \pi} \int_\opoint (\psi_t \rmd \vartheta - \psi_p \rmd \varphi) \\
&=& ( n \psitz - m \psipz )|_{res} + 
( n \psi_t^{m,n} - m \psi_p^{m,n})|_{res} \, e^{i u_\opoint} + c.c.\nonumber
\end{eqnarray}
where all radial functions must be evaluated on the rational surface defined by $\iota=n/m$, even if not explicitly stated in the following notation. 
Introducing the helical flux function
\be
\label{psih}
\psi_h(\psitz, u)=  m \psi_p - n \psi_t \, ,
\ee
$\so$ can be written in terms of this flux through the $\opoint$-point orbit as 
\begin{eqnarray}
-\so &=& \psi_h(\psitz, u_\opoint)  \nonumber \\
&=& \psi_h^0 + \left|\psi_h^{m,n}\right| \, e^{i (u_\opoint+\alpha_h^{m,n})}  + c.c.
\end{eqnarray}
where $\psi_h^0=\psi_h^0(\psitz)$ is the unperturbed equilibrium flux, whereas $\left| \psi_h^{m,n} \right|$ and $\alpha_h^{m,n}$ are the amplitude and phase of the $(m,n)$ Fourier component.

The calculation of the $\sx$ flux follows the same steps, with $u_\opoint$ being substituted by $u_\xpoint$, and remembering that $u_\xpoint$ is shifted by $\pi$ with respect to $u_\opoint$: $u_\opoint+\alpha_h^{m,n}=0, \pi$ and $u_\xpoint+\alpha_h^{m,n}=\pi,0$, depending on the sign of the magnetic shear.
This yields:
\begin{eqnarray}
-\so(\psitz) &=& \psi_h^0 + 2 \left| \psi_h^{m,n} \right| \cos(u_\opoint+\alpha_h^{m,n}) \nonumber \\
\label{So_psih}
&=& \psi_h^0 \pm 2 \left| \psi_h^{m,n} \right|\\
-\sx(\psitz) &=& \psi_h^0 + 2 \left| \psi_h^{m,n} \right| \cos(u_\xpoint+\alpha_h^{m,n})  \nonumber\\
\label{Sx_psih}
&=& \psi_h^0 \mp 2 \left| \psi_h^{m,n} \right| \\
\label{So_Sx}
\phiox &\equiv& \so-\sx = \mp 4 \, \left| \psi_h^{m,n} \right|
\end{eqnarray}
with all radial functions evaluated on the rational surface. The minus sign in equation \eqref{So_Sx} corresponds to a negative magnetic shear at the rational surface ($\rmd \iota/\rmd \psitz < 0$), which implies $u_\opoint+\alpha_h^{m,n}=0$, while the positive sign corresponds to the opposite case ($\rmd \iota/\rmd \psitz > 0$), with $u_\opoint+\alpha_h^{m,n}=\pi$.

From an operative point of view, the $\phiox$ flux can be computed both from equation \eqref{So_Sx}, or solving numerically the line integrals in equation \eqref{actionB_fl} for $\gamma=\opoint$ and $\gamma=\xpoint$. In the first case one needs to evaluate the helical flux perturbation at the resonant surface; in the second case, one needs to know the path of the island extrema.

The amplitude of the magnetic island can then be computed from the formula (similar to equation (90) of \citet{escande24})
\be
W_{\phiox} = 4 \, \sqrt{\left| \frac{\phiox}{2 m \frac{ \rmd \iota}{\rmd \psitz}} \right|} \left(\frac{\rmd r}{\rmd \psitz}\right),
\label{Deltar}
\ee
where the factor $(\rmd \psitz/ \rmd r)^{-1}$ brings the width in units of a lengths. It implies a clear relation between the unperturbed flux $\psitz$ and a radial coordinate $r$ in meters: when $r$ is the radius of circular flux surfaces cross-section of the zeroth-order equilibrium, the factor $(\rmd \psitz/ \rmd r)^{-1}$ considers the same amplitude at any poloidal or toroidal cuts. All radial functions, as the magnetic shear or the $(\rmd \psitz/ \rmd r)^{-1}$ term, must be evaluated at the resonant surfaces.


\section{Magnetic islands as new equilibrium systems}
\label{sec:sheq}

\begin{figure}
\centering
\includegraphics[width=0.5\columnwidth]{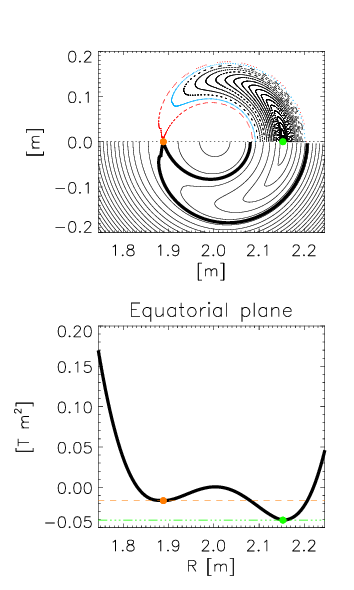}
\caption{Top. $\phis=0$ discretized field map of a ($1,7$) island (top half of the section) and corresponding flux surface contour (bottom half) in a circular RFP. Red and light-blue curves in the half top panel represent the two last flux surfaces from the Flit code \citep{flit}, the thick line in the bottom half panel the separatrix computed by the Sheq code \citep{martines11}, and coloured dots the $\xpoint$ (orange) and $\opoint$ (green) points. Bottom: the helical flux $\psi_h$ on the equatorial plane passing through the $\xpoint$ (orange) and $\opoint$ (green) points.}
\label{fig:horizontalflux}
\end{figure}

In section \ref{sec:b} we stated the equivalence between the magnetic action along a helical path $\gamma$ and the helical flux. From this perspective, equations \eqref{So_psih}-\eqref{Sx_psih} can be written in a shorter notation:
\begin{eqnarray}
\label{So_c}
- \so &=& \psi_h|_\opoint \\
\label{Sx_c}
- \sx &=& \psi_h|_\xpoint
\end{eqnarray}
and therefore
\be
\label{SoSx_c}
\phiox \equiv \so-\sx = - \psi_h|_\opoint + \psi_h|_\xpoint
\ee
where $\psi_h|_\opoint$ means the helical flux evaluated along the line defined by the $\opoint$-point, and similarly $\psi_h|_\xpoint$. From a geometrical point of view, $\so$ can be interpreted as the helical flux $\psi_h|_\opoint$ through the the surface delimited by the orbit of the center of the island ($\opoint$ point), whereas $\sx$ as the helical flux $\psi_h|_\xpoint$ through the edge of the island (the orbit defined by the $\xpoint$ point). In this section, as well as in section \ref{sec:inside}, a way to compute the magnetic fluxes through any island flux surface is shown.

The perturbative approach in section \ref{sec:b} assumes an integrable unperturbed magnetic field configuration, i.e., the equation of motion can be solved to give non-chaotic magnetic field lines and therefore conserved magnetic flux surfaces. In presence of general magnetic perturbations the system is not integrable, and flux surfaces are destroyed.
Apart from an axisymmentric system, there is only one other known integrable system, i.e. that of helical symmetry, that defines conserved magnetic flux surfaces, analogous to the constant energy surfaces. In both cases, the Hamiltonian is time-independent, and the equivalence $t=\varphi$ holds.

Equations \eqref{psit_sh}-\eqref{psip_sh}, adding to the unperturbed equilibrium a single Fourier perturbation, define a helical integrable Hamiltonian. To integrate it, we make use of the change of coordinates $(\psitz,\vartheta,\phis)\mapsto(\psitz,u,\phis)$, where $u=m \vartheta - n \varphi$ is the helical variable. This change of coordinates defines a new time-independent Hamiltonian: the helical flux $\psi_h(\psitz,u)= m \psi_p-n \psi_t$, that can be assumed as radial variable of any system with a helical symmetry \citep{hazeltine1992}. In fact, in the new coordinates, the identifications with the $(p,q,t)$ variables are: $q=x^2=u$, $t=x^3=\varphi$ and therefore $p=\psi_t$, $H=\psi_h$.

The island domain can be modeled as a helical equilibrium configuration, and a reconstruction of such equilibria has been implemented in a code named SHEq \citep{martines11}, now extended to the tokamak case too. The method is based on the superposition of an axisymmetric equilibrium and of a first-order helical perturbation computed according to Newcomb's equation supplemented with edge magnetic field measurements \citep{zanca04}; more details are in section \ref{sec:c}. The helical flux contours give the shape of the flux surfaces of a helical domain. An example of such surfaces is shown in figure \ref{fig:horizontalflux} (bottom half of the top panel) for a magnetic island in a RFP plasma. The accuracy of the flux surface reconstruction is confirmed by the corresponding discretized field map (top half) obtained with the field line tracing code Flit, which integrates the field lines with the same helical Fourier perturbations \citep{flit}. The value of helical flux through the surfaces delimited by the $\xpoint$ and $\opoint$ point orbits (their intersection with the poloidal plane respectively plotted in figure with an orange and green dot) can be identified from the value of the helical flux at the extrema of its profile on the equatorial plane (figure \ref{fig:horizontalflux}, bottom panel). The values of the helical flux above the value at the separatrix (orange dashed line in figure \ref{fig:horizontalflux} bottom) label the external flux surfaces with respect to the magnetic island, so the helical flux profile computed by the SHEq code is here cutted at the separatrix to restrict the computation to the island domain.
The values of $\psi_h$ in the well between the $\xpoint$ and the $\opoint$ point correspond to the island flux surfaces; whereas the values of $\psi_h$ in the other well correspond to the circular flux surfaces around the axisymmetric equilibrium axis.
It is worth noting that the evaluation of the helical flux on the $\xpoint$ and $\opoint$ points permits to evaluate the $\phiox$ flux from equation \eqref{SoSx_c}, and therefore the island width from equation \eqref{Deltar}.

We can now define a new reference frame having its axis on the $\opoint$-point of the island. From here on we introduce the new index $\cdot_\subsheq$ to explicitly identify the quantities related to the helical magnetic flux surfaces $\Sigma_\subsheq$ from the $\opoint$-point to the separatrix.
A time-independent canonical change of coordinates allows us to write the Hamiltonian of the helical system, $\psi_h$, in its action-angle form, $\rhos$, where both the poloidal and toroidal fluxes across the helical flux surfaces are constants of the motion \citep{momo11} and therefore functions of $\rhos$. Due to the time-independence of the canonical tranformation, $\psi_h$ and $\rhos$ define the same flux surfaces.
Using the definition of the canonical action coordinate, the identifications between canonical and magnetic coordinates, and the perturbed fluxes in equations \eqref{psit_sh}-\eqref{psip_sh}, the toroidal flux through $\Sigma_\subsheq$ is defined by:
\begin{eqnarray}
\label{fh_action}
\torfluxs(\rhos) &=& \frac{1}{2 \pi} \oint p(E,q) \, \rmd q = I(E) \\
\label{fh}
&=& \frac{1}{2 \pi} \oint_{\Sigma_\subsheq} \psi_t(\psi_h,u) \, \rmd u\,.
\end{eqnarray}
Equation \eqref{fh_action} is the standard formula for the canonical action coordinate, usually indicated with the symbol $I(E)$ for a given energy value $E$, and $q$ the canonical position \citep{arnold2013}. In equation \eqref{fh} the identifications with canonical coordinates have been used, the constant energy value $E$ corresponding to the constant value of $\psi_h$ on $\Sigma_\subsheq$, and the expression \eqref{psih} for the helical flux has been inverted to obtain $\psi_t(\psi_h,u)$. It is worth noting that on the right hand side of this equation the perturbed quantities appear, as $\psi_t$ and $\psi_h$, while on the left hand side we have the quantities through the island flux surfaces, identified by $\cdot_\subsheq$. As a side remark, the action coordinate in equation \eqref{fh_action} is not related to the magnetic action in equation \eqref{actionB}. The angle coordinate, defined on the helical axis, is defined by \citep{arnold2013}
\begin{eqnarray}
\label{angle_formula}
\us &=& \int_{q_0}^q \frac{\partial p(I,q')}{\partial I} \, \rmd q' \\
\label{angle_uh}
&=&  \int_0^u \frac{\partial \psi_t(\rhos,u')}{\partial \torfluxs} \, \rmd u' \, ,
\end{eqnarray}
that turns out to be the straight-helical-like angle defined on the helical axis (the $\opoint$-point of the island) which increases by $2\pi$ one turn around every helical flux surface. 

Equation \eqref{fh} implies that both magnetic fluxes (the action $\torfluxs$ and the Hamiltonian $\rhos$ of the system) are constant of the motion, i.e. are constant on magnetic flux surfaces. Moreover, magnetic field lines written in the action-angle coordinates are straight lines in the ($\us, \phis$) plane. 
The definitions of a helical angle and of a helical flux, 
\begin{eqnarray}
\label{uh_thetah}
\us &=& m \, \thetas - n \, \phis \\
\label{psih_psiph}
\rhos &=& m \, \polfluxs-n \,\torfluxs \, ,
\end{eqnarray}
bring to the definition of the new poloidal-like angle $\thetas$ defined on the island $\opoint$ point and, implicitly, of the poloidal flux through $\Sigma_\subsheq$, $\polfluxs=(\rhos+n \torfluxs )/m$. The rotational transform related to straight-field-lines in the plane ($\thetas, \phis$) is defined by
\be
\label{iota_7}
\iotas=\frac{\rmd \thetas}{\rmd \phis} = \frac{\rmd \polfluxs}{\rmd \torfluxs} 
\ee
which counts the poloidal and toroidal turns of an island magnetic field line seen by an external observer. Moreover, the Jacobian and the metric elements of the $(\rhos,\thetas,\phis)$ coordinate system allow us to calculate any geometric quantity related to the island domain, as the island volume, using the procedure that will be described in the next section for the stand-alone coordinate system.

We remark that the helical fluxes, $\psi_h(\psitz,u)$ and $\rhos(\torfluxs)$ in equations \eqref{psih} and \eqref{psih_psiph}, respectively, are the same flux, due to the fact that the coordinate transformation $(\psitz,u,\phis)\mapsto(\rhos,\thetas,\phis)$ is equivalent to a time-independent canonical transformation, that does not change the Hamiltonian of the system. They differ just by the value of $\psi_h(\psitz,u)$ on the helical axis, which ensures that $\rhos$, $\torfluxs$ and $\polfluxs$ vanish there:
\be
\rhos=\psi_h(\psitz,u)-\psi_h|_\opoint
\label{psiHpsih}
\ee
being $\psi_h|_\opoint=\psi_h(\psitz,u)|_\opoint$, whereas $\rhos|_\opoint=0$. Equation \eqref{psiHpsih}, together with equations \eqref{So_c} and \eqref{Sx_c} for $\so$ and $\sx$, permits the interpretation of $\phiox$ (which is the magnetic flux through the ribbon given by the $\opoint$ and $\xpoint$ point orbits, as depicted in figure \ref{fig:nastroDE}) as the helical flux through the separatrix surface. In fact,
\begin{eqnarray}
- \so &=& \psi_h|_\opoint \\
- \sx &=& \psi_h|_\xpoint = \psi_h|_\opoint + \rhos|_\xpoint
\end{eqnarray}
simply yield
\be
\phiox= \so-\sx = -\rhos|_\xpoint\,.
\label{phiox_Sx}
\ee
Note that these considerations also apply to the next section.


\section{Magnetic islands as stand-alone domains}
\label{sec:inside}

\begin{figure}
\centering
\includegraphics[width=0.5\columnwidth]{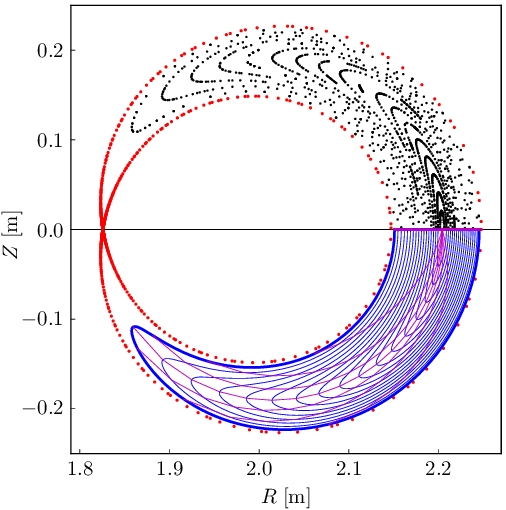}
\caption{$\varphi=0$ discretized field map of a (1,1) island in a circular tokamak (top half of the section) and corresponding flux-coordinate grid (bottom half) with the $\rhoi=$ const lines (in blue, thick for $\rhoi=1$) and the $\thetai=k\,\pi/8$, $k$ integer lines (in purple, thick for $\thetai=0,\pi$).}
\label{fig:inside_distance}
\end{figure}

Magnetic islands, even if embedded in a global toroidal magnetic field, can be regarded as separated plasma domains. In a previous work \citep{predebon18} we described a method to characterize geometrically every isolated domain from discretized field maps. These maps are usually the outcome of a field-line tracing code or an MHD code: starting from a set of points in the usual cylindrical coordinates $(R,Z,\varphi)$ -- with $R$ the distance from the axis of the torus, $Z$ the distance from the equatorial plane, and $\varphi$ the geometrical toroidal angle -- the method allows to obtain a magnetic coordinate system $(\torfluxi,\thetai,\phii)$ in the island, with the following expression for the magnetic field $\mathbf{B}$,
\be
\label{Bcan}
\mathbf{B}=\frac{1}{2\pi}\left(\nabla\torfluxi\times\nabla\thetai + \iotai(\torfluxi)\nabla\phii\times\nabla\torfluxi\right),
\ee
where $\iotai(\torfluxi)=\rmd\polfluxi/\rmd\torfluxi$ is the rotational transform inside the island, $\phii$ the toroidal angle, and $\thetai$ the poloidal angle such that the field lines are straight on the $(\thetai,\phii)$ plane, with $d\thetai/d\phii=\iotai(\torfluxi)$. As $\phii$ is the geometric toroidal angle, these coordinates are called symmetry flux coordinates \citep{dha}. We remark that the $(\torfluxi,\thetai,\phii)$ and $(\torfluxs,\thetas,\phis)$ systems, being flux coordinates sharing the same toroidal angle, are mathematically equivalent. The indexes $\cdot_\subitalo$ and $\cdot_\subsheq$ denote the different procedure used to generate the respective flux coordinate systems.

Due to the high curvature of the surfaces in the proximity of the $\xpoint$-point, the method developed by \citet{predebon18} does not allow to get an accurate description of the geometry in that region, thus we limit our reconstruction to the surface immediately preceding the separatrix. This is assumed to be the boundary of our domain.

There is freedom in the choice of the radial coordinate. Once a normalized radial coordinate $\rhoi$ is defined such that $\rhoi=0$ on the magnetic axis and $\rhoi=1$ on the last closed magnetic surface of the domain, and the Jacobian matrix $d(R,Z,\varphi)/d(\rhoi,\thetai,\phii)$, or equivalently  $d(x,y,z)/d(\rhoi,\thetai,\phii)$, is known, we can calculate the (inverse) metric tensor $g^{ij}=\nabla x^i\cdot\nabla x^j$ and the Jacobian $J=\sqrt{g}$, which for the coordinates $(\rhoi,\thetai,\phii)$ will be explicitly written as $\jaci$.

Being the metric tensor well defined in the whole island domain, we introduce here the observables that we intend to compare with the other approaches. Let us consider the width of the island itself. This can be measured with the ruler or can be calculated using the metric tensor. At a given toroidal angle $\phii=\bar\phii$, for a fixed the poloidal angle $\thetai=\bar\thetai$, the (curvilinear) distance covered along the $\rhoi$ direction from the magnetic axis to a generic surface with $\rhoi\leq 1$ is
\be
L|_{(\bar\thetai,\bar\phii)}(\rhoi)=\int_0^\rhoi g_{\rhoi\rhoi}^{1/2}(\rhoi',\bar\thetai,\bar\phii)\,\rmd\rhoi',
\ee
where we have used the infinitesimal line element expression $\rmd l^2=g_{ij}\,\rmd x^i\,\rmd x^j$ restricted to the radial direction. In figure \ref{fig:inside_distance}, we show the $\phii=0$ section of a $(1,1)$ island in a circular tokamak based on the RFX-mod geometry, obtained again with the field line tracing code Flit. At this section the $\thetai=0$ and $\thetai=\pi$ coordinate lines correspond to the horizontal cut of the island, so that the island width is simply given by
\be
\label{eq:width_italo}
\widthi|_{\bar\phii=0} = L|_{(0,0)}(1) + L|_{(\pi,0)}(1)
\ee
Other useful quantities for a comparison with the other approaches include the volume enclosed by the surface with radius $\rhoi$
\be
\label{eq:vol_italo}
\voli(\rhoi) = \int_{[0,\rhoi]\times[0,2\pi]\times[0,2\pi]} \jaci\,\rmd\rhoi'\,\rmd\thetai\,\rmd\phii,
\ee
as well as the poloidal (the ribbon poloidal flux, as is called in \citet{dha}) flux
\be
\label{eq:psip_italo}
\polfluxi(\rhoi) = \frac{1}{2\pi}\int_{[0,\rhoi]\times[0,2\pi]\times[0,2\pi]} B^{\thetai}\,\jaci \,\rmd\rhoi'\,\rmd\thetai\,\rmd\phii,
\ee
and the toroidal flux
\be
\label{eq:psit_italo}
\torfluxi(\rhoi) = \frac{1}{2\pi}\int_{[0,\rhoi]\times[0,2\pi]\times[0,2\pi]} B^{\phii}\,\jaci \,\rmd \rhoi'\,\rmd\thetai\,\rmd\phii
\ee
where the poloidal and toroidal contravariant components of the field are given by $B^\thetai=\mathbf{B}\cdot\nabla\thetai$ and $B^\phii=\mathbf{B}\cdot\nabla\phii$, respectively.

Combining equations \eqref{eq:psip_italo}-\eqref{eq:psit_italo} to define the helical flux $\helfluxi = m \polfluxi - n \torfluxi$, the island width can be again inferred from equations \eqref{Deltar} with $\phiox=-\helfluxi|_\xpoint$, as in equation \eqref{phiox_Sx}, since the helical flux on the island axis vanishes identically. In this case, the $\phiox$ flux is calculated at $\rhoi=1$, not at the separatrix, as a line integral following one of the tips of the island, which is the curve that best approximates the separatrix, corresponding to the angle $u_\subitalo=m\thetai-n\phii=\pm\pi/2$.


\section{A comparison of the different approaches}
\label{sec:c}

\begin{figure}
\centering
\includegraphics[width=0.7\columnwidth]{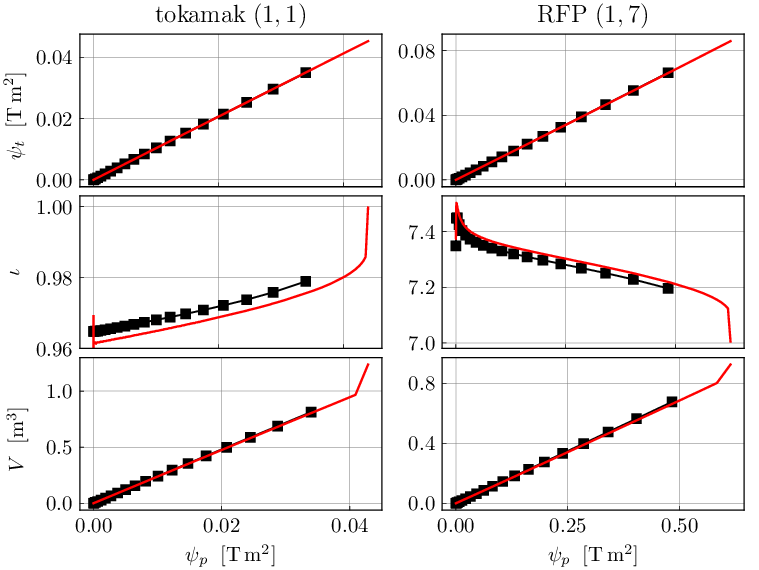}
\caption{Toroidal flux (top panels), $\iota$ (mid panels), and volume (bottom panels) as a function of the poloidal flux for a (1,1) tokamak island (left column) and a (1,7) RFP island (right column), for the approaches of section \ref{sec:sheq} (red lines) and section \ref{sec:inside} (black).}
\label{fig:flussi}
\end{figure}

\begin{table*}
\hspace*{-5mm}
\begin{tabular*}{\linewidth}{cc|cccccccc}
  \hline
  \hline
  island & surface & $\vols$\,[m$^3$]$^\dag$  & $\torfluxs$\,[Tm$^2$]$^\dag$  & $\polfluxs$\,[Tm$^2$]$^\dag$ & $\voli$\,[m$^3$]$^\ddag$ & $\torfluxi$\,[Tm$^2$]$^\ddag$  & $\polfluxi$\,[T\,m$^2$]$^\ddag$ \\
  \hline
  \multirow{2}{*}{tok\,$(1,1)$} & $\rhoi=1$   & 0.806  & 3.49$\times 10^{-2}$  &  3.38$\times 10^{-2}$ & 0.813  & 3.50$\times 10^{-2}$ &  3.40$\times 10^{-2}$ \\
                                & separ.      & 1.237  & 4.50$\times 10^{-2}$  &  4.39$\times 10^{-2}$ &    --  &   --                &    --                 \\
  \multirow{2}{*}{RFP\,$(1,7)$} & $\rhoi=1$   & 0.673  & 6.61$\times 10^{-2}$  &  4.83$\times 10^{-1}$ & 0.676  & 6.63$\times 10^{-2}$ &  4.83$\times 10^{-1}$ \\
                                & separ.      & 0.925  & 8.58$\times 10^{-2}$  &  6.25$\times 10^{-1}$ &    --  &   --                &    --                 \\
  \hline
  \hline
  &  & $\phiox$\,[T\,m$^2$]$^\dag$ & $W_{\phiox}$\,[cm]$^\dag$ & $\phiox$\,[T\,m$^2$]$^\ddag$ & $W_{\phiox}$\,[cm]$^\ddag$ & $\phiox$\,[T\,m$^2$]$^\circ$ & $W_{\phiox}$\,[cm]$^\circ$ \\
  \hline
  \multirow{2}{*}{tok\,$(1,1)$} & $\rhoi=1$  & 1.12$\times 10^{-3}$  & 9.96   & 1.02$\times 10^{-3}$  & 9.49   & --                    & --     \\
                                & separ.     & 1.34$\times 10^{-3}$  & 10.86  & --                   & --     &  1.35$\times 10^{-3}$  & 10.93  \\
  \multirow{2}{*}{RFP\,$(1,7)$} & $\rhoi=1$  & 2.05$\times 10^{-2}$  & 12.01  & 1.90$\times 10^{-2}$  & 11.57  & --                    & --     \\
                                & separ.     & 2.40$\times 10^{-2}$  & 13.01  & --                   & --     &  2.40$\times 10^{-2}$  & 13.12  \\
  \hline
  \hline
  &  &  $\widthi|_{\bar\varphi=0}$\,[cm]$^\ddag$ & $W_\mathrm{r}|_{\bar\varphi=0}$\,[cm]  \\
  \hline
  \multirow{2}{*}{tok\,$(1,1)$} & $\rhoi=1$  & 9.45  &  9.45  \\
                                & separ.     &  --   & 10.07  \\
  \multirow{2}{*}{RFP\,$(1,7)$} & $\rhoi=1$  & 10.99 & 10.97  \\
                                & separ.     & --    & 11.86  \\
  \hline
  \hline
\end{tabular*}
\caption{For a (1,1) tokamak island and a (1,7) RFP island, $\cdot^\dag$ quantities refer to the perturbative method of section \ref{sec:sheq}, $\cdot^\ddag$ quantities to the non-perturbative method of section \ref{sec:inside}, $\cdot^\circ$ quantities to the perturbative method of section \ref{sec:b}; in the second row $W_{\phiox}$ is calculated from $\phiox$ by means of equation \eqref{Deltar} for the 3 different methods: $\phiox$\,$^\dag=m\polfluxs-n\torfluxs$, $\phiox$\,$^\ddag=m\polfluxi-n\torfluxi$, $\phiox$\,$^\circ$ as defined in equation \eqref{So_Sx}. $W_\subitalo|_{\bar\varphi=0}$ is the width as defined in equation \eqref{eq:width_italo}. $W_\mathrm{r}|_{\bar\varphi=0}$ is the width measured with the ruler at $\varphi=0$.}
\label{tab:island}
\end{table*}


In the following, we provide a comparison of some observables using the different approaches described above. For this comparison, we consider the islands already introduced in the previous sections, namely a (1,1) island in a circular tokamak and a (1,7) island in the RFP, both based on RFX-mod, a circular toroidal device with major radius $R_0=2$ m and minor radius $a=0.46$ m which can perform operation in either configuration \citep{sonato2003}.

The reconstruction of the magnetic island topology is based on the calculation of the helical perturbations to the zeroth-order equilibrium. According to the procedure developed by \citet{zanca04}, this is done solving a Newcomb-like equation, i.e. the linearized force-balance equation with the linearized Ampere's law ensuring a divergence-free magnetic field, using the external magnetic measurements as boundary conditions. Due to the set of measurements available in RFX-mod (48 toroidal arrays of 4 poloidally equispaced probes for both the radial and toroidal field components), a rich spectrum of Fourier components can be reconstructed for the magnetic perturbations. In this work we focus on the dominant resonant Fourier component generating the $(m,n)$ island, as specified in equations \eqref{psit_sh}-\eqref{psip_sh}. The mode is $(m,n)=(1,1)$ for the tokamak case, pulse 38818 at t=362 ms, and $(m,n)=(1,7)$ for the RFP case, which is the averaged discharge described in \citet{momo20}. The resulting islands have been characterized geometrically in \citet{predebon18}, where the discretized field maps have been produced by the field line tracing code Flit \citep{flit} which is indeed based on the Newcomb's perturbed fluxes as input.

As already mentioned, the comparison with the stand-alone approach of section \ref{sec:inside} is possible only within the surface $\rhoi=1$, which for both the RFP and tokamak islands is the last surface of the Poincar\'e section before the separatrix. For the other approaches the radial domain extends from the $\opoint$-point of the island to the separatrix.

In figure \ref{fig:flussi} and table \ref{tab:island} we summarize the results of the comparison. In the figure, the toroidal flux, the rotational transform and the volume profiles are plotted as a function of the poloidal flux for the approaches of section \ref{sec:sheq} (red lines) and section \ref{sec:inside} (black). In particular, the poloidal and toroidal fluxes and the rotational transform defining the island flux surfaces are computed from equations \eqref{fh} and \eqref{iota_7} in the approach of section \ref{sec:sheq}, and from equations \eqref{eq:psip_italo}-\eqref{eq:psit_italo} in the approach of section \ref{sec:inside}. The volume is computed in both cases from the definition \eqref{eq:vol_italo}, using the related coordinate system and Jacobian. The comparison can be considered satisfactory. The small difference in the recontruction of the rotational transform profile (around 0.4\% at the surface $\rhoi=1$ for the tokamak, 0.5\% for the RFP), explains the pointwise differences in the fluxes that appear in the following table. 

For the two islands, in the table we review the most relevant quantities calculated at the $\xpoint$-point and at the last surface at the Poincar\'e map ($\rhoi=1$). For the perturbative method of section \ref{sec:sheq} applied to the $\rhoi=1$ surface, we assume $\polfluxs|_{\rhoi=1}=\polfluxi|_{\rhoi=1}$ and derive the other quantities based on this reference value. The island volume, the toroidal and poloidal fluxes through the island are reported in the first block of the table. In the second block, the $\phiox$ flux is computed from equation \eqref{So_Sx} for the approach of section \ref{sec:b}, as the helical flux through the separatrix using equation \eqref{phiox_Sx} for the approach of section \ref{sec:sheq}, and as the path integral along the (best approximation of the) $\xpoint$-point identified by the angle $u_\subitalo=m\thetai-n\phii=\pm\pi/2$ in the geometric approach of section \ref{sec:inside} (proving also that $\phiox=-\sx$ when the fluxes vanish on the $\opoint$-point). Then, for the 3 methods, the related island width is calculated with equation \eqref{Deltar}. To complete the table, we also report the island width as resulting from equation \eqref{eq:width_italo} and as measured with a ruler from the Poincar\'e maps at $\varphi=0$, $W_\mathrm{r}|_{\bar\varphi=0}$.

The method which best estimates $W_\mathrm{r}|_{\bar\varphi=0}$ is that of section \ref{sec:inside}, which is not surprising as the island geometry is directly derived from the Poincar\'e map itself: the $\widthi|_{\bar\varphi=0}$ width from equation \eqref{eq:width_italo} perfectly matches $W_\mathrm{r}|_{\bar\varphi=0}$. From the $\phiox$ flux calculated as path integral along one of the tips of the island, we provide the width $W_{\phiox}$ from equation \eqref{Deltar}, which yields a value within $1\%$ and $5\%$ error with respect to $W_\mathrm{r}|_{\bar\varphi=0}$, for tokamak and RFP respectively.

On the other hand, the perturbative methods of sections \ref{sec:b} and \ref{sec:sheq} overestimate the total island width, within $6\%$ and $10\%$ error with respect to $W_\mathrm{r}|_{\bar\varphi=0}$, for tokamak and RFP case respectively. We recall that the formula of equation \eqref{Deltar} for $W_{\phiox}$ comes from the formal analogy between the typical text-book deformation of the flux surfaces around the resonance due to the opening of a magnetic island and the phase diagram of a pendulum, valid for small perturbations. This explains the error introduced by the application of this formula to the specific cases shown here, where the perturbation cannot be considered small. Large perturbations cause a displacement of the $\opoint$-point and $\xpoint$-point from the rational surface where the radial derivatives are calculated. However, the correction to $\phiox$ when calculated as the helical flux through the island separatrix (method of section \ref{sec:sheq}), with respect to its estimate based on the axisymmetric equilibrium (method of section \ref{sec:b}), slightly improves the evaluation of the width.


\section{Summary}

We have considered three different approaches for a geometric characterization of magnetic islands, (i) the first one perturbative, (ii) the second one perturbative but leading to the definition of a new single-helicity equilibrium, (iii) the third one non-perturbative based on the availability of a discretized field map.

In the first perturbative method the island is described in a Hamiltonian context using the definition of action $S_\gamma$ of a magnetic system. The novelty of this approach, first introduced by \citet{escande24}, is the definition of a coordinate-independent flux with a clear geometrical meaning -- here named $\phiox$ -- related to the island width through equation \eqref{Deltar}. This formula, coming from the analogy between a textbook magnetic island and the eye-of-cut shape of the phase diagram of a pendulum, is very similar to the classical formulas for the island width \citep{hazeltine1992, park08}, but with a new meaning of the flux that appears in all other formulations. The second method, developed by \citet{martines11} and \citet{momo11} for the RFP equilibrium and here extended to the tokamak configuration, starts from a perturbative approach that sums the zeroth-order equilibrium and the single resonant perturbation generating the island, then leading to the definition of a new single-helicity equilibrium using definitions specific to Hamiltonian mechanics. The third, non-perturbative method, is based on the availability of a discretized field map and defines the island domain through its geometrical definitions, as first developed by \citet{predebon18}. 

A detailed comparison has been carried out among the three methods applying them to two experimental islands of RFX-mod, both tokamak and RFP. We have provided an estimate of some observables showing that they are, in general, in good agreement among each other. As a novelty, moreover, we have extended the use of the expression \eqref{Deltar} for the island width in a broader context than the perturbative approach in which it was first developed, thanks to a more comprehensive geometric interpretation of the $\phiox$ flux. In particular, the $\phiox$ flux, originally defined as the flux through the ribbon defined by the $\opoint$ and $\xpoint$ point orbits, is identified as the helical flux through the island separatrix.

As a final remark, the (third) non-perturbative method, strictly based on the geometry of the flux surfaces, is the one which provides the most accurate description of the island observables for a large part of the island domain, failing, however, to describe the separatrix due to the high curvature of the flux surfaces in the neighborhood of the $\xpoint$-point. This approach applies to any isolated region of the plasma, without any assumptions on the symmetry of the system and the possible interactions with other perturbations with different helicities. On the other hand, the two perturbative methods provide a geometric characterization which is in reasonable agreement with the non-perturbative method, at least in the cases with a strong helical symmetry. Easier to apply, based as they are on a single-mode perturbation from a linear Newcomb-like analysis, these methods can satisfactorily highlight the most relevant features of an island without the need to build discretized field maps and/or use MHD codes, providing a viable method for a fast description of a magnetic island domain.


\section*{Acknowledgements}

We are grateful to D.~F. Escande and P. Zanca for reading the manuscript and providing useful observations.


\bibliographystyle{jpp}
\bibliography{text}


\end{document}